\documentclass[conference]{IEEEtran}
\usepackage{graphicx}
\usepackage{amssymb,amsmath,bm}
\usepackage{textcomp}

\usepackage{multirow}
\usepackage{array}

\newcolumntype{L}[1]{>{\raggedright\let\newline\\\arraybackslash\hspace{0pt}}m{#1}}
\newcolumntype{C}[1]{>{\centering\let\newline\\\arraybackslash\hspace{0pt}}m{#1}}
\newcolumntype{R}[1]{>{\raggedleft\let\newline\\\arraybackslash\hspace{0pt}}m{#1}}

\ifCLASSINFOpdf
\else
\fi

\begin{document}
\title{Investigation of Synthetic Speech Detection Using Frame- and Segment-Specific Importance Weighting}

\author{\IEEEauthorblockN{Ali Khodabakhsh, Cenk Demiro\u{g}lu}
\IEEEauthorblockA{Electrical and Computer Engineering Department\\
\"{O}zye\u{g}in University, 34794, Istanbul, Turkey\\
Email: ali.khodabakhsh@ozu.edu.tr, cenk.demiroglu@ozyegin.edu.tr}
%\and
%\IEEEauthorblockN{Mustafa Caner \"{O}zbay}
%\IEEEauthorblockA{Institute\\
%Vestel, Istanbul, Turkey\\
%Email: mustafa.ozbay@vestel.com.tr}
}
% \and
% \IEEEauthorblockN{James Kirk\\ and Montgomery Scott}
% \IEEEauthorblockA{Starfleet Academy\\
% San Francisco, California 96678--2391\\
% Telephone: (800) 555--1212\\


%Telephone: (800) 555--1212, Fax: (888) 555--1212}

% \IEEEspecialpapernotice{(Invited Paper)}

\maketitle

\begin{abstract}
Speaker verification systems are vulnerable to spoofing attacks which presents a major problem in their real-life deployment.
To date, most of the proposed synthetic speech detectors (SSDs) have weighted the importance of different segments of speech equally. However, different attack methods have different strengths and
weaknesses and the traces that they leave may be short or long term acoustic artifacts. 
Moreover, those may occur for only particular phonemes or sounds.
Here, we propose three algorithms that weigh likelihood-ratio scores of individual frames, phonemes, and sound-classes depending on their importance for the SSD.
Significant improvement over the baseline system has been obtained for 
known attack methods that were used in training the SSDs.
However, improvement with unknown attack types was not substantial.
Thus, the type of distortions that were caused by the unknown systems were different and could not be captured better with the proposed SSD compared to the baseline SSD.
\end{abstract}


% \fi
\IEEEpeerreviewmaketitle

\section{Introduction}

One of the biggest obstacles in deployment of speaker verification technology in real-life scenarios, especially in high-security applications such as telephone banking, is the difficulty in countering spoofing attacks.
Even though verification of speaker identity through human voice has been shown to be successful~\cite{greenberg20132012}, state-of-the art verification systems have been shown to be vulnerable to spoofing attacks using speech synthesis and voice conversion~\cite{wu2015spoofing}. 

Most of the literature on the spoofing problem has focused on algorithms that were designed to counter specific types of attacks.
For example, one method of synthesizing speech is the HMM-based approach where smooth speech parameters are generated and speech is synthesized with a vocoder.
Even though HMM-based synthesis can successfully spoof the modern verification systems, it is also easy to detect by exploiting the unnaturally smooth trajectories of the parameters~\cite{satoh2001robust}\cite{tomoki2007speech}\cite{chen2010speaker}. 
Moreover, because the vocoder typically has a minimum-phase filter, phase was also used for detecting HMM-based synthesis since natural speech spectrum is not minimum phase~\cite{de2012evaluation}.

Even though unit selection synthesis is relatively harder to detect, it is also challenging to deploy in the context of spoofing since unlike the HMM-based approach that can adapt to the target with seconds of data, unit selection requires hours of training data.
Such large amounts of data is hard to collect for each target speaker in most practical cases.
Existing synthetic speech detectors (SSDs) typically use jumps in fundamental frequency at the concatenation points for detection~\cite{ogihara2005discrimination}\cite{de2012synthetic}.

Voice conversion algorithms can also be used for spoofing~\cite{wu2015spoofing}.
Because they typically use minimum-phase vocoders, phase was used in~\cite{wu2012detecting}\cite{wu2012study} for detecting voice-converted speech.
Moreover, some voice conversion systems exhibit low parameter variability across an utterance compared to natural speech and that was also exploited for detecting voice conversion~\cite{alegre2013spoofing}.

There are also SSDs that are independent of the attack type.
One promising approach is to use a local binary pattern (LBP) analysis for feature extraction~\cite{alegre2013one}.
In that approach, a one-class classifier is trained with features derived only from natural speech.
The classifier learns the spectro-temporal model of speech and can detect synthetic signals that do not fit well to that model.
In~\cite{zhizheng2015information}, i-vectors are used both for speaker verification and synthetic speech detection. The detector and speaker verification scores are fused to make a final decision.

Here, we investigate several detectors without attack-specific prior assumptions. 
Our approach is based on the hypothesis that long- and/or short-duration artifacts will be observed in the synthetic speech without any constraints on the type of artifacts.
Artifacts that typically occur in stop sounds during synthesis because of their rapidly changing dynamics and sudden glitches that occur frequently with the unit selection systems are examples of short-duration artifacts.
Overly-smooth parameters generated with HMM-based synthesis is an example to long-duration artifacts.
The SSD algorithm should be sensitive to both types of artifacts to be effective.

In this paper, we have investigated SSDs that can capture both short and long-duration artifacts.
The first SSD was developed using an unsupervised approach where a Gaussian mixture model (GMM) is trained for natural speech and a GMM is trained for synthetic speech.
After aligning each speech frame with a Gaussian, each Gaussian component is treated as an independent detector and detector scores are fused with logistic regression. 

Our second method is based on designing detectors that are focused on detecting artifacts in specific phonemes.
This approach can be successful at detecting phoneme-specific artifacts in synthetic speech.
However, some of the phonemes are not observed frequently enough in most utterances.
To reduce the data sparsity issue, broad-level sound class detectors are used in a third approach.
Similar to the Gaussian approach, score fusion is done for the phoneme- and class-based methods.

All three methods performed substantially better than the baseline detector that treats all Gaussians and phonemes equally for the known attack types.
However, the proposed systems did not substantially improve the baseline system for unknown attack types.
Fusing the three proposed detectors further improved the SSD performance both in known and unknown conditions.

\section{Synthetic Speech Detectors}

\begin{figure}
	\centering
	\includegraphics[width=\linewidth]{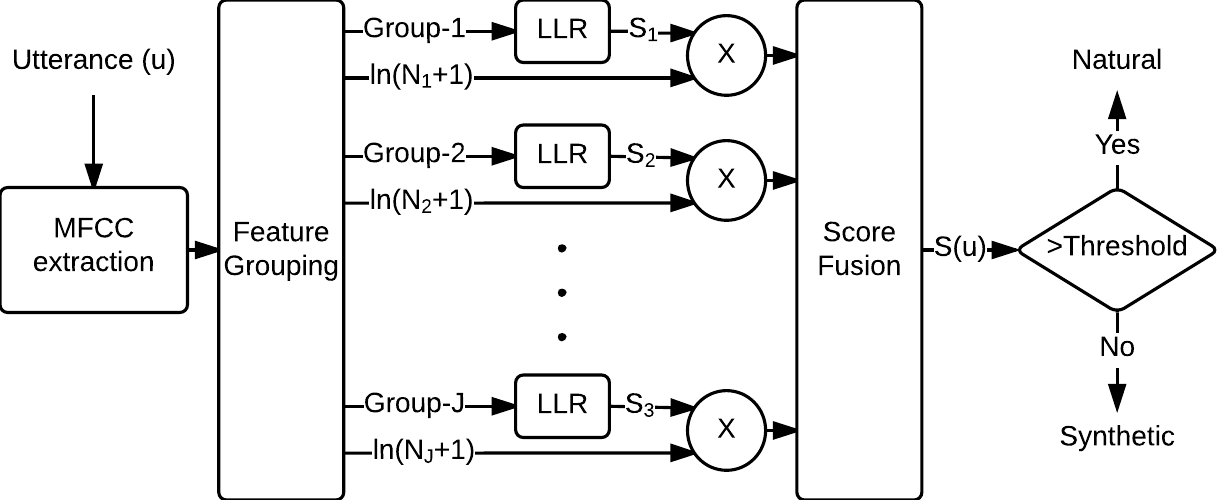}
	\caption{Overview of the proposed synthetic speech detectors.}
	\label{fig:Diagram}
\end{figure}

An overview of the proposed system is shown in Fig.~\ref{fig:Diagram}.
Mel-frequency cepstral coefficients (MFCC) are first extracted from the speech utterance.
Then, the feature vectors are grouped together into $J$ groups.
In one approach, vectors that are aligned with the same Gaussian component of a GMM are grouped together.
In another approach, feature vectors that belong to the same phoneme or sound class constitute a group.
Details of grouping are described in the next section.

After grouping, log-likelihood ratio (LLR) detection is done for each group of feature vectors.
To compute LLR, a GMM is trained for natural speech and a GMM is trained for synthetic speech.
Same GMMs are used for all $J$ groups.
Once the score of each group is computed, score fusion is done using a logistic regression function to compute the final score $S(u)$.
A hard threshold is used to compute the final decision.

In the baseline detector, which does not use any grouping, given an utterance $u$, assuming independent speech frames
\begin{equation}
LLR(u)=\frac{1}{N}\sum_{i=1}^{N} log({\bf x}_i|{\bf \Lambda}_{nat}) - log({\bf x}_i|{\bf \Lambda}_{syn}),
\label{Eq:simple}
\end{equation}
where $N$ is the total number of frames, $x_i$ is the feature vector for the $i^{th}$ frame, $\bf{\Lambda_{nat}}$ is the canonical model of GMM for the natural speech, and $\bf{\Lambda_{syn}}$ is the canonical model of GMM for the synthetic speech.
The final decision is done using a hard threshold for $LLR(u)$.

In the proposed approach, the decision is based on the utterance score
\begin{equation}
S(u)=\varPhi(S_1, S_2, ..., S_J)
\end{equation}
where $\varPhi$ is a nonlinear function and score $S_j$ for each group $j$ is
\begin{equation}
\frac{1}{N_j}\sum_{i=1}^{N_j} log({\bf x}_i^{(j)}|{\bf \Lambda}_{nat}) - log({\bf x}_i^{(j)}|{\bf \Lambda}_{syn}).
\label{Eq:group}
\end{equation}

The rationale of this approach is to develop detectors that are focused on different segments of speech and weigh each segment depending on its information content.
For example, nasals are typically not modeled well by vocoders because of the spectral dip in nasals that are not modeled with an all-pole model.
A detector that is focused only on nasals can detect those artifacts.
Similarly, synthetic speech may contain some short-duration glitches that are not observed in natural speech.
Even though those artifacts may be detectable by some of the Gaussian components in synthetic GMMs.
when the frame likelihoods are averaged as in Eq.~\ref{Eq:simple}, those short-duration events may not be detected because of the low weight they get and noise introduced in other frames. 
Focusing on those highly informative Gaussians regardless of their durations and assigning them high weight can improve the detection performance in those cases.

\subsection{Duration-based Weighting}

Distribution of the frame-level LLR values approximately follow a Gaussian distribution in most utterances.
By averaging the LLR scores, as done in Eq.~\ref{Eq:group}, assuming Gaussianity, a maximum-likelihood (ML) estimate of the mean is found.
Considering the fact that the ML estimate of the mean of a Gaussian has an estimation variance that is inversely proportional with the number of observations, reliability of the detector $j$ increases when $N_j$ increases.
To take the estimation variance, hence the uncertainty of the detector scores, into account, we propose the duration-weighted score
\begin{equation}
S_j^{'} = \ln(N_j+1)S_j
\end{equation}
where $\ln(.)$ is the natural logarithm.

\section{Feature Grouping Methods}

Three feature grouping strategies are investigated.
In the phoneme-based approach, each phoneme constitutes a group.
Thus, feature vectors that occur within a particular phoneme type in the utterance are grouped together. 

One of the problems with the phoneme-based approach is that some of the utterances provided in the challenge were short (~2-3seconds) which means that many of the phonemes were not observed in those cases.
Because broad acoustic-phonetic sound classes share similar acoustic properties, we hypothesized that if a system performs poorly in synthesizing a phoneme, it will most likely perform poorly for the other phonemes that are acoustically similar.
Thus, to make more data available for each group, a class-based approach is used for grouping in the second approach.
In the class-based approach, five sound classes are used: vowels, nasals, glides, stops, and rest.
The rest class contains all phonemes that do not belong to the other four classes. 

The phoneme- and class-based methods are good at detecting artifacts that occur in relatively long segments.
However, they are not designed for detecting sudden glitches that can easily occur with unit selection systems or some of the voice conversion systems.
Location of those glitches are random for the most part and they may not be detected with detectors that are focused on long-duration segments. 

To address the issue of short-duration artifact detection, we propose Gaussian-based grouping where each frame in the utterance is first aligned with the GMM of natural speech.
Then, frames that are aligned with the same Gaussian are grouped together.
This approach allows detection of frame-level artifacts and assign them high weight even though they may occur infrequently in the utterance.

\section{Experiments}

\subsection{Experiment Setup}
The synthetic speech detectors were trained with 19 MFCCs together with the delta and delta-delta features.
In short-time analysis, frame length was 25msec and frame rate was 10msec.
Bigaussian voice activity detection (VAD) was used where energy of the speech and noise frames are modeled with single Gaussians and likelihood ratio detector is used to detect speech frames.

The baseline synthetic speech detector had a 512-component GMM to model natural speech.
Similarly, synthetic speech was modeled with 512-component GMM.
For natural speech, GMM training was initialized using k-means clustering.
The GMM for synthetic speech was adapted from the GMM of the natural speech using a maximum a posteriori (MAP) approach.
Experiments with synthetic speech GMM that was trained independent of the natural speech GMM were also performed for comparison.

The phoneme-based approach requires a phoneme recognizer since the transcriptions of the challenge data were not available.
The Hungarian phoneme recognizer \cite{schwarz2006phoneme} was trained with WSJ-CAM database and used here for phoneme recognition.
A total of 37 phonemes were used.
Outputs of the phoneme recognizer were mapped to sound classes and used in sound-class based detector also.

The spoofing challenge database\footnote{http://www.spoofingchallenge.org/asvSpoof.pdf} was used for training, development and evaluation of all systems.
The BOSARIS toolkit \cite{brummer2013bosaris} was used to train the logistic regression algorithm that was used for fusing the scores of detectors.
	
\subsection{Results and Discussion}

\begin{table*}[ht]
\caption{Performance of the Baseline and Proposed Detectors in Terms of Equal-error-rates (EERs) for the Development and Evaluation Data.
Results are Presented With and Without Duration-weighting.
S1, S2, and S5 Systems Use Voice Conversion (VC).
S3 and S4 Systems Use HMM-based Synthesis.
Best Performing Algorithm for Each Attack Type is Shown in Bold.}
\label{tab:system_results}
\centering
% \vspace{2mm}
\begin{tabular}{ll|r|r|r|r|r|r|r|r|r|r|}
\cline{3-12}
                                                   &              & \multicolumn{6}{c|}{Normal}                                                                                                                                                         & \multicolumn{4}{c|}{Duration-based weighted}                                                                              \\ \cline{3-12} 
                                                   &              & \multicolumn{2}{c|}{LLR}                                  & \multicolumn{4}{c|}{Logistic Regression}                                                                                & \multicolumn{4}{c|}{Logistic Regression}                                                                                  \\ \cline{3-12} 
                                                   &              & \multicolumn{1}{c|}{noAdapt} & \multicolumn{1}{c|}{Adapt} & \multicolumn{1}{c|}{Class} & \multicolumn{1}{c|}{Phoneme} & \multicolumn{1}{c|}{Gaussian} & \multicolumn{1}{c|}{Fusion} & \multicolumn{1}{c|}{Class} & \multicolumn{1}{c|}{Phoneme} & \multicolumn{1}{c|}{Gaussian} & \multicolumn{1}{c|}{Fusion}   \\ \hline
\multicolumn{1}{|l|}{\multirow{6}{*}{Development}} & S1           & 0.47                         & 0.76                       & 0.68                       & 0.69                         & 0.47                          & \textbf{0.41}               & 0.54                       & 0.54                         & 0.51                          & 0.46                          \\ \cline{2-12} 
\multicolumn{1}{|l|}{}                             & S2           & 10.24                        & 5.12                       & 3.37                       & 3.41                         & 1.89                          & \textbf{1.83}               & 2.99                       & 3.13                         & 2.26                          & 2.20                          \\ \cline{2-12} 
\multicolumn{1}{|l|}{}                             & S3           & 0.07                         & 0.07                       & \textbf{0.03}              & 0.09                         & 0.20                          & 0.17                        & \textbf{0.03}              & 0.09                         & 0.18                          & 0.11                          \\ \cline{2-12} 
\multicolumn{1}{|l|}{}                             & S4           & 0.04                         & 0.09                       & 0.05                       & \textbf{0.03}                & 0.25                          & 0.20                        & \textbf{0.03}              & 0.07                         & 0.20                          & 0.13                          \\ \cline{2-12} 
\multicolumn{1}{|l|}{}                             & S5           & 4.63                         & 3.04                       & 2.78                       & 2.86                         & 1.72                          & 1.57                        & 2.65                       & 2.72                         & 1.59                          & \textbf{1.47}                 \\ \cline{2-12} 
\multicolumn{1}{|l|}{}                             & Total        & 4.21                         & 2.42                       & 1.92                       & 1.77                         & 1.17                          & \textbf{1.11}               & 1.67                       & 1.66                         & 1.19                          & 1.14                          \\ \hline
\multicolumn{1}{|l|}{\multirow{13}{*}{Evaluation}} & S1           & 0.54                         & 0.57                       & 0.55                       & 0.56                         & 0.48                          & \textbf{0.41}               & 0.51                       & 0.50                         & 0.46                          & 0.42                          \\ \cline{2-12}
\multicolumn{1}{|l|}{                            } & S2           & 9.24                         & 4.47                       & 2.78                       & 2.71                         & 1.89                          & \textbf{1.75}               & 2.63                       & 2.44                         & 2.15                          & 2.03                          \\ \cline{2-12}
\multicolumn{1}{|l|}{                            } & S3           & 0.07                         & \textbf{0.02}              & 0.04                       & 0.04                         & 0.18                          & 0.12                        & \textbf{0.02}              & 0.03                         & 0.15                          & 0.09                          \\ \cline{2-12}
\multicolumn{1}{|l|}{                            } & S4           & 0.07                         & \textbf{0.03}              & 0.05                       & 0.05                         & 0.17                          & 0.11                        & \textbf{0.03}              & 0.04                         & 0.13                          & 0.08                          \\ \cline{2-12}
\multicolumn{1}{|l|}{                            } & S5           & 3.95                         & 1.72                       & 1.99                       & 2.14                         & 1.48                          & \textbf{1.36}               & 1.89                       & 1.97                         & 1.50                          & 1.40                          \\ \cline{2-12}
\multicolumn{1}{|l|}{                            } & S6           & 3.49                         & 1.35                       & 1.39                       & 1.40                         & 1.09                          & \textbf{0.98}               & 1.31                       & 1.24                         & 1.13                          & 1.01                          \\ \cline{2-12}
\multicolumn{1}{|l|}{                            } & S7           & 1.91                         & 1.65                       & 0.84                       & 0.87                         & 0.75                          & \textbf{0.63}               & 0.85                       & 0.94                         & 0.70                          & 0.65                          \\ \cline{2-12}
\multicolumn{1}{|l|}{                            } & S8           & \textbf{0.46}                & 1.03                       & 0.76                       & 0.85                         & 0.83                          & 0.70                        & 0.71                       & 0.87                         & 0.79                          & 0.70                          \\ \cline{2-12}
\multicolumn{1}{|l|}{                            } & S9           & \textbf{0.43}                & 1.26                       & 0.93                       & 1.02                         & 0.76                          & 0.65                        & 0.94                       & 1.02                         & 0.68                          & 0.64                          \\ \cline{2-12}
\multicolumn{1}{|l|}{                            } & S10          & \textbf{27.24}               & 29.62                      & 32.14                      & 33.59                        & 30.05                         & 29.81                       & 31.39                      & 32.25                        & 29.88                         & 29.78                         \\ \cline{2-12}
\multicolumn{1}{|l|}{                            } & Known        & 2.77                         & 1.36                       & 1.08                       & 1.10                         & 0.84                          & \textbf{0.75}               & 1.02                       & 0.99                         & 0.88                          & 0.81                          \\ \cline{2-12}
\multicolumn{1}{|l|}{                            } & Unknown      & 6.70                         & 6.98                       & 7.21                       & 7.54                         & 6.70                          & \textbf{6.55}               & 7.04                       & 7.27                         & 6.64                          & \textbf{6.55}                 \\ \cline{2-12}
\multicolumn{1}{|l|}{                            } & All          & 4.74                         & 4.17                       & 4.15                       & 4.32                         & 3.77                          & \textbf{3.65}               & 4.03                       & 4.13                         & 3.76                          & 3.68                          \\ \hline
\end{tabular}
% \vspace{-3mm}
\end{table*}

Experimental results for the development and evaluation data are shown in Table~\ref{tab:system_results}.
The baseline LLR detector is trained with two different methods.
In one approach (LLR-noAdapt), two independent GMMs are trained for the natural and synthetic speech.
In the second approach (LLR-Adapt), a GMM is trained for natural speech and then adapted to the synthetic speech using MAP adaptation.

The LLR-Adapt system performed better for known conditions while LLR-noAdapt performed better for unknown conditions.
Thus, even though LLR-Adapt performed better than LLR-noAdapt on average, it could not generalize as good as the LLR-noAdapt.
This result indicates that, during GMM training, some of the novel clusters in the synthetic data that were  useful for ambiguity detection, could not be modeled well with adaptation of GMM for natural speech. 

Gaussian-based system performed better than class- and phoneme-based methods both for known and unknown conditions.
In particular, Gaussian-based approach performed better for the S1, S2, and S5 methods, all of which are voice conversion algorithms.
Unlike the phoneme- and class-based systems, Gaussian-based detector can learn to detect short-duration artifacts.
Thus, the presence of short-duration acoustic distortions seems to be more informative for detecting voice conversion attacks. 

Class-based system performed better for S3 and phoneme-based system performed better for S4 attack methods.
Both S3 and S4 are generated with HMM-based TTS.
Unlike the voice conversion systems, HMM-based TTS systems generate smooth trajectories.
Thus, sudden acoustic distortions are rarely generated with those systems.
In this case, overly-smooth longer segments seem to be more informative for detection.
Small distortions in a long segment can be detected well with class- and phoneme-specific detectors that are focused on particular segments.
However, Gaussian-based approach is not expected to be as successful with this type of attack because speech frames are generated with a maximum-likelihood approach in HMM-based synthesis.
Thus, the parameter generation algorithm is designed to generate high likelihoods for each frame and individual Gaussians are not expected to detect the artifacts in features.

Duration-based weighting consistently improved class- and phoneme-based performance.
However, for the Gaussian-based approach, performance improved slightly for the unknown systems and degraded slightly for the known systems.
We believe there are at least two major factors behind this result.
Firstly, because an important strength of the Gaussian-approach is its ability to detect short-time artifacts, weighting with duration can hurt its performance.
Secondly, duration of observed Gaussians can change significantly depending on the spoofing system used which can increase the variability of features and make the detection task harder.
Because ASR systems take phoneme durations into account during recognition, that effect is not as important in the phoneme- and class-based methods.

\begin{figure}
	\centering
	\includegraphics[width=\linewidth, trim=25 5 35 20, clip]{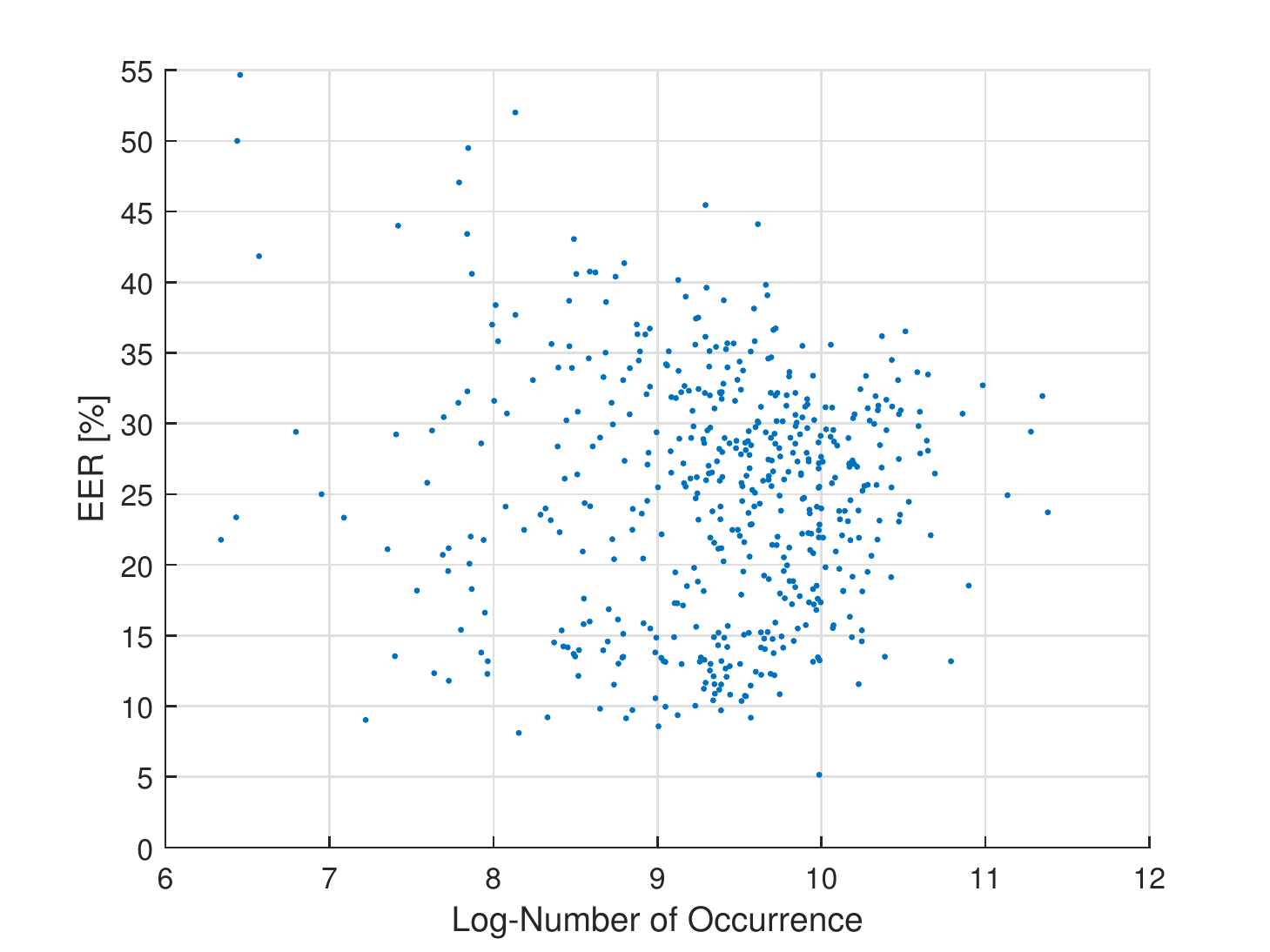}
	\caption{Detection performance of each Gaussian component versus its logarithm of number of occurrence in the development utterances is shown.}
	\label{fig:GaussianScatter}
\end{figure}

\begin{figure}
	\centering
	\includegraphics[width=\linewidth, trim=25 5 35 20, clip]{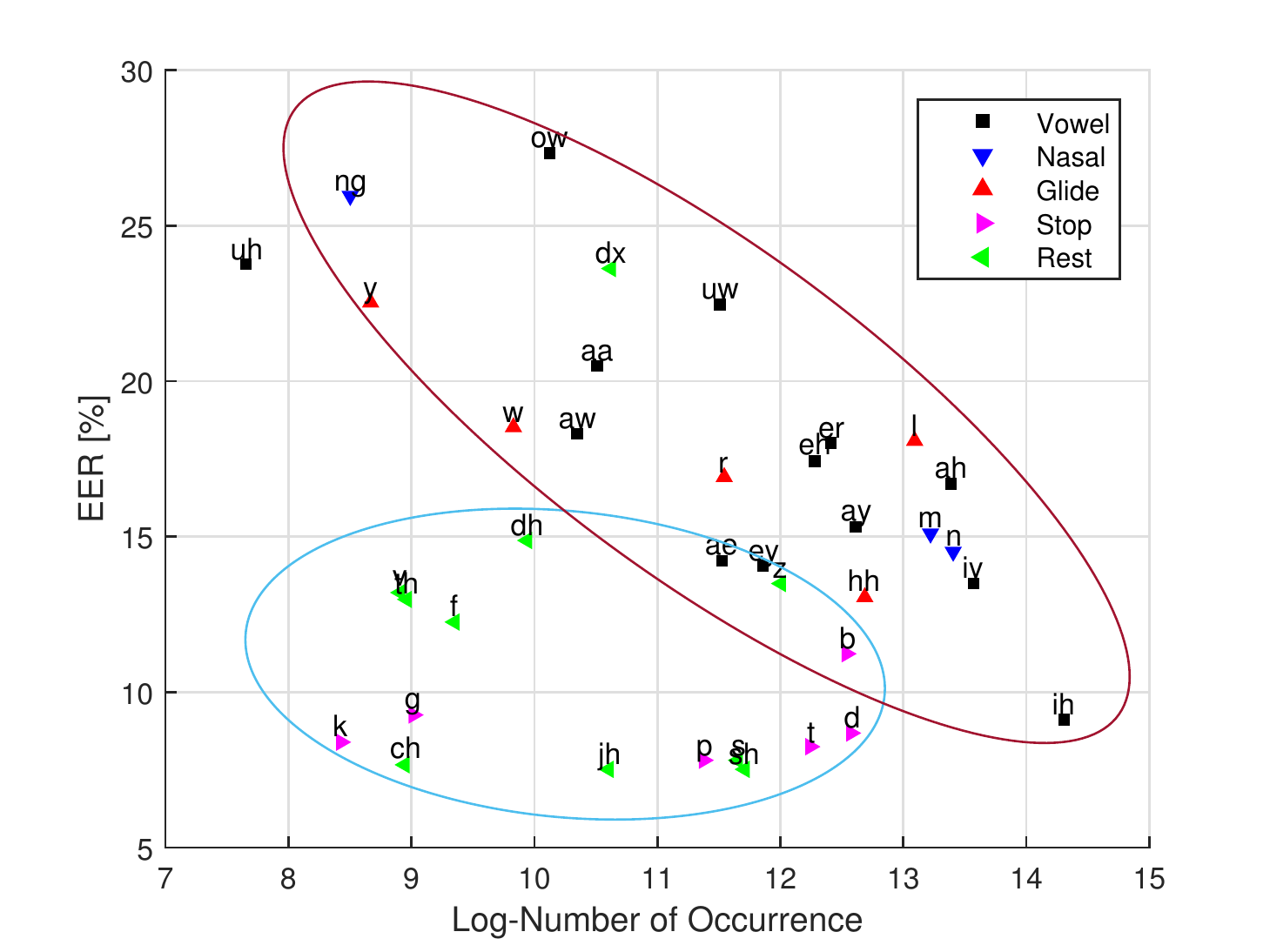}
	\caption{Detection performance of each phoneme versus its logarithm of number of occurrence in the development utterances is shown. Phonemes that are in the same sound-class are shown with the same color and shape.}
	\label{fig:PhonemeScatter}
\end{figure}

The core hypothesis in the proposed system was that different Gaussians, phonemes, sound-classes contribute different amounts of information for synthetic speech detection.
To test that hypothesis, experiments were performed with each Gaussian, phoneme, and sound-class separately.
For the Gaussian case, results are shown in Fig.~\ref{fig:GaussianScatter}, for the phoneme case, results are shown in Fig.~\ref{fig:PhonemeScatter}.
In both cases, large variation in detection performance can be observed which verifies our hypothesis. 

\begin{table}
% \vspace{-2mm}
\caption{Performance of Each of the Sound-class Detectors Measured in Terms of Equal-error-rates (EERs) for the Development Data.
Frequency of Observation in Development Utterances is also Shown for Each Class Type.}
\label{tab:classresults}
\centering
% \vspace{2mm}
\begin{tabular}{|L{0.80cm}|R{0.75cm}|R{0.75cm}|R{0.75cm}|R{0.75cm}|R{0.75cm}|R{0.75cm}|R{0.75cm}|}
\hline
Class & \multicolumn{1}{c|}{S1} & \multicolumn{1}{c|}{S2} & \multicolumn{1}{c|}{S3} & \multicolumn{1}{c|}{S4} & \multicolumn{1}{c|}{S5} & \multicolumn{1}{c|}{All} & \multicolumn{1}{c|}{Freq.} \\ \hline
Vowel & 3.52                    & 13.30                   & \textbf{0.65}           & \textbf{0.74}           & 7.26                    & 6.35                     & 0.542                      \\ \hline
Nasal & 8.90                    & 20.82                   & 5.09                    & 5.86                    & 13.79                   & 11.62                    & 0.156                      \\ \hline
Glide & 9.33                    & 21.69                   & 4.10                    & 4.44                    & 15.92                   & 12.15                    & 0.118                      \\ \hline
Stop  & \textbf{2.24}           & \textbf{4.78}           & 0.70                    & 0.78                    & \textbf{6.77}           & \textbf{3.68}            & 0.112                      \\ \hline
Rest  & 8.97                    & 10.58                   & 3.32                    & 3.78                    & 16.08                   & 9.43                     & 0.072                      \\ \hline
\end{tabular}
% \vspace{-3mm}
\end{table}

Detector results for the class-based system is shown in Table~\ref{tab:classresults}.
Performance of each class is significantly different from each other and they change substantially depending on the attack method.
Also note that, even though vowel class is observed more than other classes, their performance is better than other systems only for HMM-based TTS attacks.
For the voice-conversion attacks, short-duration stop sounds become more informative even though they occur far less frequently than the vowels.

Fig.~\ref{fig:GaussianScatter} shows the correlation of number of occurrences vs EER computed with each of the 512 Gaussians.
Even though EER and durations have a negative correlation, the pattern is weak and does not impact the overall detector performance significantly.
This result is inline with the finding that duration-based weighting does not improve the performance of the Gaussian-based system.

The effect of duration is more significant with phoneme-based detector compared to the Gaussian-based detector.
Duration versus EER is shown in Fig.~\ref{fig:PhonemeScatter} where a stronger negative correlation is observed compared to the Gaussian case especially for the vocalic sounds.
The correlation disappears for some of the highly informative stop and fricative sounds.

The proposed detectors performed substantially better than the baseline detectors for known attack types.
However, the difference is not substantial for the unknown attack types.
To further boost the performance, the detectors were fused with a second stage of logistic regression algorithm.
The fusion improved performance both for known and unknown attack types which indicate that the detectors generate complementary information.

\section{Conclusion and Future Work}

We have investigated a multi-detector approach for synthetic speech detection where each detector is focused on a particular acoustic segment.
The Gaussian-based detector performed better in voice conversion attacks.
Phoneme- and class-based detectors performed better for HMM-based synthesis attacks.
Duration-based feature normalization improved the phoneme- and class-based systems but not the Gaussian-based system.
The proposed systems performed substantially better than the baseline system in known attack types.
In unknown attacks, the improvement was not substantial.
Fusing the scores of proposed detectors further improved the performance in both known and unknown conditions.

Our goal in this paper was to take a commonly used likelihood ratio based SSD and use it in a segment-specific manner.
The hypothesis here was that different segments contribute different amounts of information and their scores should be weighted accordingly.
Results confirmed our hypothesis.
Because we did not assume any prior information, we have used the commonly used MFCC features.
In the future work, we will investigate a richer set of features and other classifiers such as SVM to further improve the detection performance.

\section*{Acknowledgment}
This work was supported by TUBITAK 1001 grant No 112E160.

\newpage
\bibliographystyle{IEEEtran}
\bibliography{mybib}

\begin{thebibliography}{10}
\providecommand{\url}[1]{#1}
\csname url@samestyle\endcsname
\providecommand{\newblock}{\relax}
\providecommand{\bibinfo}[2]{#2}
\providecommand{\BIBentrySTDinterwordspacing}{\spaceskip=0pt\relax}
\providecommand{\BIBentryALTinterwordstretchfactor}{4}
\providecommand{\BIBentryALTinterwordspacing}{\spaceskip=\fontdimen2\font plus
\BIBentryALTinterwordstretchfactor\fontdimen3\font minus
  \fontdimen4\font\relax}
\providecommand{\BIBforeignlanguage}[2]{{%
\expandafter\ifx\csname l@#1\endcsname\relax
\typeout{** WARNING: IEEEtran.bst: No hyphenation pattern has been}%
\typeout{** loaded for the language `#1'. Using the pattern for}%
\typeout{** the default language instead.}%
\else
\language=\csname l@#1\endcsname
\fi
#2}}
\providecommand{\BIBdecl}{\relax}
\BIBdecl

\bibitem{greenberg20132012}
C.~S. Greenberg, V.~M. Stanford, A.~F. Martin, M.~Yadagiri, G.~R. Doddington,
  J.~J. Godfrey, and J.~Hernandez-Cordero, ``The 2012 nist speaker recognition
  evaluation.'' in \emph{INTERSPEECH}, 2013, pp. 1971--1975.

\bibitem{wu2015spoofing}
Z.~Wu, N.~Evans, T.~Kinnunen, J.~Yamagishi, F.~Alegre, and H.~Li, ``Spoofing
  and countermeasures for speaker verification: a survey,'' \emph{Speech
  Communication}, vol.~66, pp. 130--153, 2015.

\bibitem{satoh2001robust}
T.~Satoh, T.~Masuko, T.~Kobayashi, and K.~Tokuda, ``A robust speaker
  verification system against imposture using an hmm-based speech synthesis
  system.'' in \emph{INTERSPEECH}, 2001, pp. 759--762.

\bibitem{tomoki2007speech}
T.~Tomoki and K.~Tokuda, ``A speech parameter generation algorithm considering
  global variance for hmm-based speech synthesis,'' \emph{IEICE TRANSACTIONS on
  Information and Systems}, vol.~90, no.~5, pp. 816--824, 2007.

\bibitem{chen2010speaker}
L.-W. Chen, W.~Guo, and L.-R. Dai, ``Speaker verification against synthetic
  speech,'' in \emph{Chinese Spoken Language Processing (ISCSLP), 2010 7th
  International Symposium on}.\hskip 1em plus 0.5em minus 0.4em\relax IEEE,
  2010, pp. 309--312.

\bibitem{de2012evaluation}
P.~L. De~Leon, M.~Pucher, J.~Yamagishi, I.~Hernaez, and I.~Saratxaga,
  ``Evaluation of speaker verification security and detection of hmm-based
  synthetic speech,'' \emph{Audio, Speech, and Language Processing, IEEE
  Transactions on}, vol.~20, no.~8, pp. 2280--2290, 2012.

\bibitem{ogihara2005discrimination}
A.~Ogihara, U.~Hitoshi, and A.~Shiozaki, ``Discrimination method of synthetic
  speech using pitch frequency against synthetic speech falsification,''
  \emph{IEICE transactions on fundamentals of electronics, communications and
  computer sciences}, vol.~88, no.~1, pp. 280--286, 2005.

\bibitem{de2012synthetic}
P.~L. De~Leon, B.~Stewart, and J.~Yamagishi, ``Synthetic speech discrimination
  using pitch pattern statistics derived from image analysis.'' in
  \emph{INTERSPEECH}, 2012.

\bibitem{wu2012detecting}
Z.~Wu, C.~E. Siong, and H.~Li, ``Detecting converted speech and natural speech
  for anti-spoofing attack in speaker recognition.'' in \emph{INTERSPEECH},
  2012.

\bibitem{wu2012study}
Z.~Wu, T.~Kinnunen, E.~S. Chng, H.~Li, and E.~Ambikairajah, ``A study on
  spoofing attack in state-of-the-art speaker verification: the telephone
  speech case,'' in \emph{Signal \& Information Processing Association Annual
  Summit and Conference (APSIPA ASC), 2012 Asia-Pacific}.\hskip 1em plus 0.5em
  minus 0.4em\relax IEEE, 2012, pp. 1--5.

\bibitem{alegre2013spoofing}
F.~Alegre, A.~Amehraye, and N.~Evans, ``Spoofing countermeasures to protect
  automatic speaker verification from voice conversion,'' in \emph{Acoustics,
  Speech and Signal Processing (ICASSP), 2013 IEEE International Conference
  on}.\hskip 1em plus 0.5em minus 0.4em\relax IEEE, 2013, pp. 3068--3072.

\bibitem{alegre2013one}
------, ``A one-class classification approach to generalised speaker
  verification spoofing countermeasures using local binary patterns,'' in
  \emph{Biometrics: Theory, Applications and Systems (BTAS), 2013 IEEE Sixth
  International Conference on}.\hskip 1em plus 0.5em minus 0.4em\relax IEEE,
  2013, pp. 1--8.

\bibitem{zhizheng2015information}
A.~Sizov, E.~Khoury, T.~Kinnunen, Z.~Wu, and S.~Marcel, ``Joint speaker
  verification and antispoofing in the $i$ -vector space,'' \emph{Information
  Forensics and Security, IEEE Transactions on}, vol.~10, no.~4, pp. 821--832,
  April 2015.

\bibitem{schwarz2006phoneme}
P.~Schwarz, P.~Matejka, L.~Burget, and O.~Glembek, ``Phoneme recognizer based
  on long temporal context,'' \emph{Speech Processing Group, Faculty of
  Information Technology, Brno University of Technology.[Online]. Available:
  http://speech. fit. vutbr. cz/en/software}, 2006.

\bibitem{brummer2013bosaris}
N.~Br{\"u}mmer and E.~de~Villiers, ``The bosaris toolkit: Theory, algorithms
  and code for surviving the new dcf,'' \emph{arXiv preprint arXiv:1304.2865},
  2013.

\end{thebibliography}

\end{document}